\documentclass[12pt,preprint]{aastex}
\def\lsim{\lower.5ex\hbox{$\; \buildrel < \over \sim \;$}}
\def\gsim{\lower.5ex\hbox{$\; \buildrel > \over \sim \;$}}

\begin{document}

\title{Evidence of Class Transitions in GRS 1915+105 from IXAE Data}

\author{Sandip K.\ Chakrabarti$^{1,2}$,  A. Nandi$^{1}$, Asit Choudhury$^3$ and Utpal Chatterjee$^3$}

\affil{$^1$S.N. Bose National Center for Basic Sciences, JD-Block, Salt Lake, Kolkata, 700098\\ India;
e-mail: chakraba@bose.res.in; anuj@bose.res.in\\
$^2$ Centre for Space Physics, Chalantika 43, Garia Station Rd., Kolkata, 700084, India; e-mail: space\_phys@vsnl.com\\
$^3$ Centre for Space Physics-Malda Branch, Atul Market, Malda, 732101, India; e-mail: space\_phys@vsnl.com\\
}

\begin{abstract}
GRS 1915+105 shows at least twelve distinct classes of light curves. 
By analysing the data obtained from Indian X-ray Astronomy Experiment (IXAE) 
instrument aboard IRS-P3 satellite, we show that in at least two
days, transitions between one class to another were observed. In these
days the so-called $\kappa$ class went to $\rho$ class 
and $\chi$ class went to $\rho$ class. In the frequency-time plane such transitions
exhibited change in quasi-periodic oscillation (QPO) frequency. We could detect that
low-frequency QPOs can occur in anticipation of a class transition several hundred 
minutes before the actual transition.
\end{abstract}


\noindent Published in Astrophysical Journal 607, 406, 2004

\section{Introduction}

GRS 1915+105 is being studied for almost ten years 
(see, Castro-Tirado et al. 1994; Mirabel \& Rodriguez, L.F. 1994 for 
early papers) quite extensively, and yet, it continues to reveal 
new physical insights into the phenomenon of black hole accretion. 
This enigmatic X-ray transient source exhibits at least twelve classes of 
light curves, if not more (Belloni et al. 2000). These twelve types have 
been designated as $\chi$, $\alpha$, $\nu$, $\beta$, $\lambda$, $\kappa$, 
$\rho$, $\mu$, $\theta$, $\delta$, $\gamma$ and $\phi$ respectively. These classes
were based on the nature of the hardness ratios exhibited in the light curve. 
In some class (e.g., $\chi$) only hard X-rays are seen, in some other classes 
(e.g., $\phi$) only soft X-rays are seen, while in the majority (such as $\kappa$,
$\lambda$, $\rho$) there were photons coming in the so-called `burst-on' and `burst-off' states 
which are similar to quick transitions between soft and hard states. Subsequently,
Nandi, Manickam \& Chakrabarti, (2000) rearranged these classes in a pattern 
more easily understandable by advective disk models (see also: Naik, Rao
\& Chakrabarti, 2002). While RXTE satellite has pointed to this object
about a thousand times, it has not been able to pinpoint exactly how 
and when a class transition actually  takes place. 

In the present {\it Letter}, we show that observations by the Pointed Proportional Counters (PPC)
in the IXAE instrument aboard IRS-P3 satellite, have revealed such 
class transitions several times. While the time-resolution and the number of energy bins of IAXE
instrument are low, the observations nevertheless are good enough to identify the 
classes and the transitions. Presently we report two such `Class-Transitions' from the
archived data of IXAE.  We also show that the source probably had `premonition' 
about the class-transition about a few hundred seconds prior to the observation 
as is evidenced by the the presence of a drift in low-frequency (LFQPO) QPO 
in the Power Density Spectrum (PDS). We observed light curves of `unknown' class just 
before the transition. We compare our results with RXTE observations in nearby times.
We present two transitions: one is $\kappa \rightarrow \rho$ transition on the 
June 22nd, 1997 and the other is $\chi \rightarrow \rho$ transition 
on the June 25th, 1997. In both the days, we find evidence of the LFQPO
in the pre- and post-transition Classes.

While the exact cause of the Class transition is not yet known, it has been 
pointed out by Nandi et al. (2000) that this is due to the variation in 
accretion rates in both the Keplerian and sub-Keplerian components. 
They actually re-arranged the overall 12 classes of Belloni (2000) into four major Classes: 
Hard Class, Soft Class, Semi-Soft Class and Intermediate Class and showed 
that the Class transitions could be understood by slow variation of 
the accretion rate of the sub-Keplerian component. Chakrabarti \& Manickam
(2000), while explaining the correlation between the duration and QPO frequency 
in burst-off states of $\kappa$ and $\lambda$ classes, pointed out that the real cause of 
QPO is most probably the oscillations of the standing shocks. Furthermore, the transition 
between the burst-off and burst-on states inside a class is most probably due to the
cooling of the outflowing wind by soft-photons from the accretion disk. As a result, two 
QPO frequencies are expected in classes which have burst-on and burst-off states. The one at a high frequency
(around $1-20$Hz or so) would be due to the shock oscillation, while the one at low frequency (say around
$0.01-0.02$Hz) would be due to the quasi-periodic cooling of the outflows.
If this is correct, then, it is possible that the basic nature of the
quasi-periodic cooling of the outflow could continue to take place 
during a class transition, while the shock itself can be weakened. This would mean that the 
low frequency QPOs could persist across class transitions. In the next Section,
we present results of our analysis and show that this is indeed the case.

In the next Section, we report the observations and show that such transitions
actually took place on the same day in the data set from a single orbit. We perform 
FFT at various times of observation and plot time-frequency diagram for 
observations in each orbit. We also compare our results with those of RXTE (which missed the
transitions) on the same date.  Finally, in \S 4, we draw our conclusions.

\section{Observational Results and Data Analysis}

The Indian X-ray Astronomy Experiment or IXAE which was lunched from Sriharikota 
Range on March 21st, 1996 and since then has carried out observations till the year 2000. 
It was flown on the Indian satellite IRS-P3, had three identical Pointed Proportional Counters (PPCs) 
(Agrawal, 1998). Each counter was filled with a gas mixture of $90$\% Argon and $10$\% Methane at a pressure of $800$ torr
with a total effective area of $1200 cm^2$.  
The operating energy range is between $2$ and $18$ keV with an average detection efficiency of $60$\% at $6$keV.
The counts are saved in the archive in two channels: in $2-6$keV and in $6-18$keV. Although the
time-resolution could be $0.1$s in Medium mode, due to restrictions in data-storage
on-board, usually observations have been carried out in $1$s resolution to have the data 
for five consecutive orbits per day. This is a great advantage which allowed one to 
observe class transitions at the cost of fast variability behaviour for which 
Rossi X-ray Timing Explorer (RXTE) is undoubtedly the champion.

In Fig. 1(a-e), we show the light curves (upper panels) and time-frequency diagrams (lower panels)
obtained in five successive orbits on June 22nd, 1997. The light curves
were obtained by adding counts from all the three PPCs from both the energy channels $2-6$ and
$6-18$keV. The lower panels were obtained by Fast Fourier Transform (FFT) 
of the photon counts in $6-18$keV. 
While in panel 1a, in the first orbit, the Class is $\kappa$, in panel 1e, the Class is $\rho$. 
In between, the light curves do not appear to belong to any known class.
The data has been analyzed by using FTOOLS and XSPEC packages of NASA. Since the
time resolution of the observation is only $1$s, we are unable to identify the high-frequency
QPOs in IXAE data. However, the LFQPOs, in burst-on/burst-off sources,
has been detected. For each panel of Figs. 1(a-e), we divide the data in segments of 
$306$s and stagger each segment by $51$ seconds. For each segment of $306$s we take the Fast Fourier
Transform (FFT) and place the PDS at its mean time. Finally, we plot contours of 
constant normalized power $P$ as a function of the frequency and time of observation in each panel. 

The times of observations have been taken from Table 1 of Yadav et al. (1999). 
There is an error of 1 day in the Date column in that Table. This we have corrected.
The observation started at 12:12:24 but Fig. 1a starts at 12:14:57, 
$153$ seconds later, because  of our choice of segment size. The observation 
ended at 19:20:50 while Fig. 1e ends at 19:18:17, $153$ seconds earlier.
There is no available record of the exact beginning and end times of each gap between orbits.
Two successive datasets are separated by $\sim 83$ minutes.
Thus, the exact time of the observation cannot be given within 2-3 minutes of accuracy. 
Only the duration is known with certainty. However, this does not affect out conclusions.

The contours in lower panel of Fig. 1a are drawn for normalized power $P=2.0,\ 2.3,\ 2.6,$
$2.9,\ 3.2,\ 3.5$ and $3.8$ (highest power is with darkest shade) clearly indicating an LFQPO frequency $\nu$
 oscillating between $log(\nu) =-1.896$ and $log(\nu)=-1.824$.
In Fig. 1b, contours are drawn for $P=1.7,\ 2,\ 2.3,\ 2.6,$ and  $2.9$. We note that in this unknown class 
the LFQPO frequency monotonically drifts from $log(\nu) = -1.937$ to  $log(\nu) = -1.793$.
In Fig. 1c, the contours are drawn for $P= 1.7,\ 2.0,\ 2.3,\ 2.6$ and $2.9$ respectively.
The drift of LFQPO frequency undergoes large amplitude fluctuation with high
reaching up to $log(\nu)\sim -1.73$. In Fig. 1d, which was drawn for  $P=1.7,\ 2.0,\ 2.3,\ 2.6,\ 2.9$ and 
$3.2$ respectively, the first failed attempt of the Class transition is
seen after around $750$s of the time the begining of the observation where a large spike is seen
in the light curve. The completely different power density spectra (PDS) in the pre- and post-transition
region produces a `mixed' PDS in the transition region, causing large scale 
noise of duration $\sim 300$s, the size of our segments over which FFTs were taken. 
In the result of the final orbit in Fig. 1e, the contours are drawn for $P= 1.7,\ 2,\ 2.5,\ 2.8,\ 3.1$ and $3.5$ respectively.
Besides the noise generated by the transition region, it is clear that the LFQPO frequency
at $log(\nu)\sim -1.75$ in the pre-transition `unknown' Class continues to drift towards higher value [$log(\nu) \sim -1.664$]
as the light curve enters into the well known $\rho$ class. Indeed, only $7$ minutes after this IXAE observation ends,
RXTE observed GRS 1915+105 for an hour (Obs. ID=20402-01-34-01) between 19:27:28 to 20:27:04 and found the
object in $\rho$ class with the same characteristics. There was an observation of RXTE on the 18th of June, 1997
(Obs. ID = 20402-01-33-00) and the class observed was $\kappa$, same as the pre-transition class in IXAE
observation on the 22nd of June, 1997 (Fig. 1a).

In Fig. 2, we show another Class transition on June 25th, 1997.  
On this day also observations were carried out for five successive orbits. The first orbital observation
begins at 11:12:15 and the fifth observation ends at 18:18:52
with gaps between the data of two successive orbits around $83$ minutes as before. 
We show only the results of the fourth orbit when actual transition took place.
The light curve (upper panel) clearly shows a transition from $\chi$ to $\rho$ Class.
The time-frequency plot after FFT of $306$s segments staggered by $51$s, shows (lower panel)
the familiar noise in PDS of duration $\sim 300$s. The Class transition  itself
took place at around $T \sim 950$s. The contours are drawn for $P=0.3,\ 0.6,\ 0.9,\ 1.2,\ 1.5,\ 1.8,\ 2.1$ 
and $2.4$ respectively. It is very clear that even from $\sim 400$s after the observation starts,
the system started showing a LFQPO with frequency $log(\nu) \sim -1.78$, while after transition,
the LFQPO drifts towards $log(\nu)\sim -1.70$. The first harmonics can also be observed
at around $log(\nu) \sim -1.4$. Assuming the gaps between the orbits are of exactly  equal
duration,  we estimate that this orbital observation ended at $\sim$ 16:36:58. 

In this case also RXTE has observed GRS 1915+105 a few minutes after this observation,
from 17:05:20 to 17:15:04 for a period of ten minutes (Obs. ID=20402-01-34-00). Fig. 3 shows the 
light curve and the PDS. The nature of the light curve matches with 
IXAE observation and the PDS shows a clear peak at $log(\nu) \sim -1.63$ with the first harmonics.
It appears that the LFQPO frequency drifted a bit in the intervening half an hour.

\section{Concluding Remarks}

In this {\it Letter}, we have presented evidences of direct transition of
GRS 1915+105 going from one class of light curve to another class of 
light curve. Three such occasions have been found in data obtained by IXAE
on June 22nd and 25th, 1997. We find that every time such a transition takes place,
the PDS in the pre- and post-transition look very different and the intervening 
period is characterized by a noise (induced due to mixing of these two
types of PDS results) and in all the cases, the LFQPO frequency 
continues through the transition. On the 22nd of June, 2003
we found that an unknown class was sandwiched by the $\kappa$ and $\rho$
class and a `failed' attempt was made for a class-transition about $90$ minutes 
prior to the actual transition. In both the dates we verified that
our results are consistent with RXTE observations also. Indeed, RXTE also 
observed on several occasions $\chi$ state followed by $\rho$ state (e.g. 
Obs IDs 20402-01-27-01 \& 20402-01-27-02 and 20402-01-29-00 \& 20402-01-30-00; see, Belloni et al. 2000) 
as in our observation of 22nd June, 1997, though exact transition time was not detected by RXTE.

This work is supported in part from a DST project (AN) and an ISRO project (SKC).
AC and UC thank Dr. Achintya Chatterjee for discussions. The authors thank Prof. P.C. Agrawal
of TIFR for making the IXAE data available to us.

{}

\vfil\eject
\begin{figure}
\vskip 0.0cm
\plotone{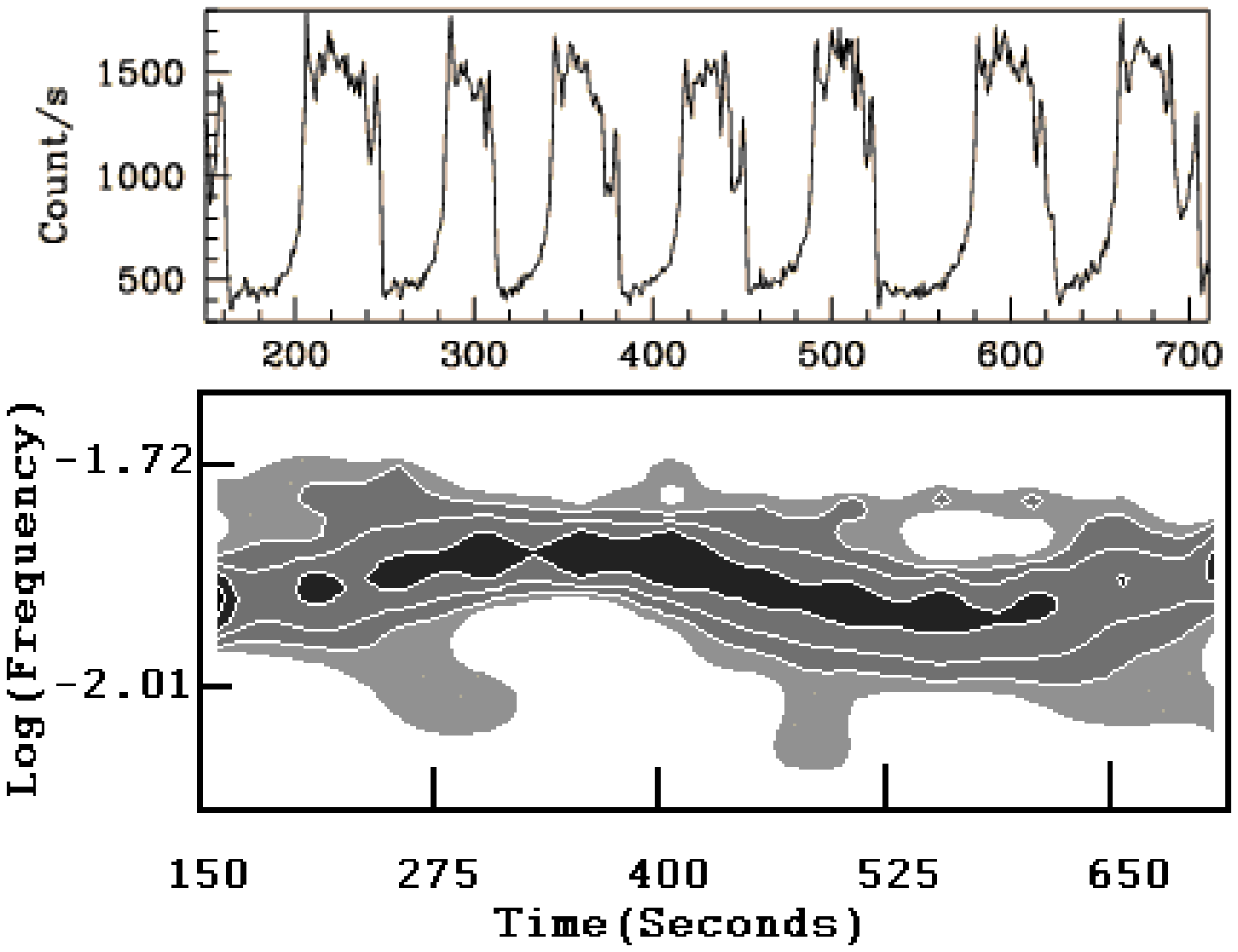}
\end{figure}

\vfil\eject
\begin{figure}
\vskip 0.0cm
\plotone{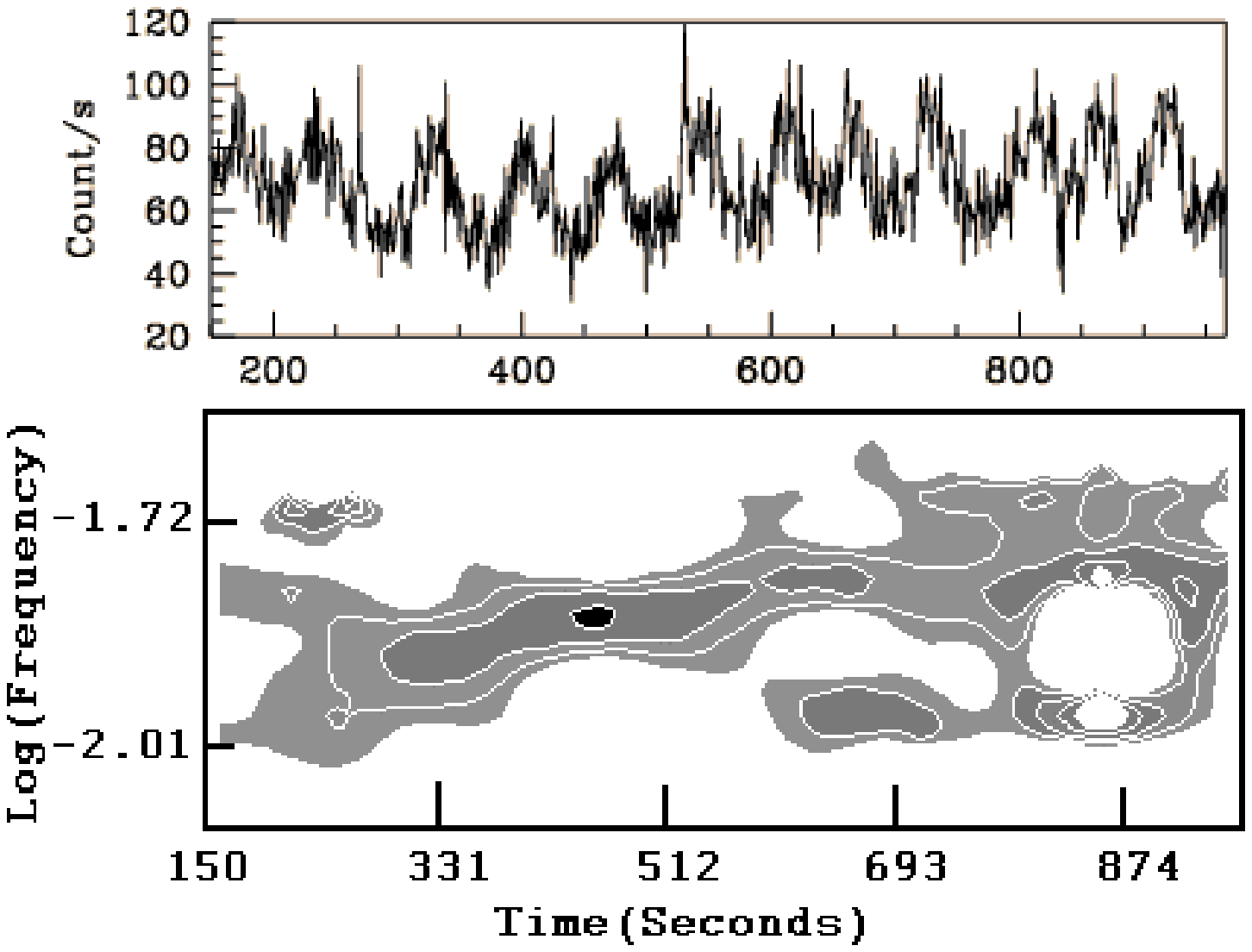}
\end{figure}

\vfil\eject
\begin{figure}
\vskip 0.0cm
\plotone{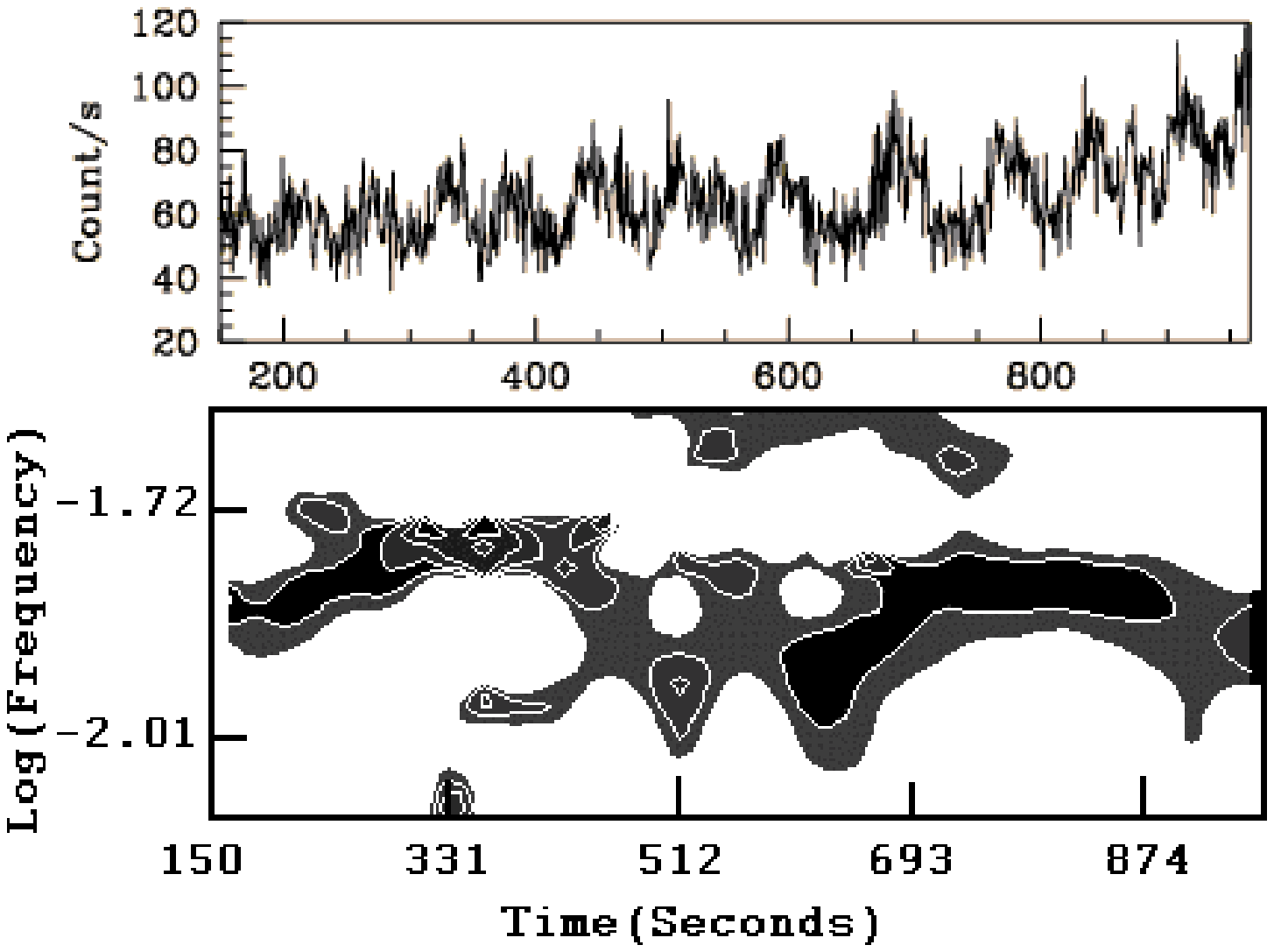}
\end{figure}

\vfil\eject
\begin{figure}
\vskip 0.0cm
\plotone{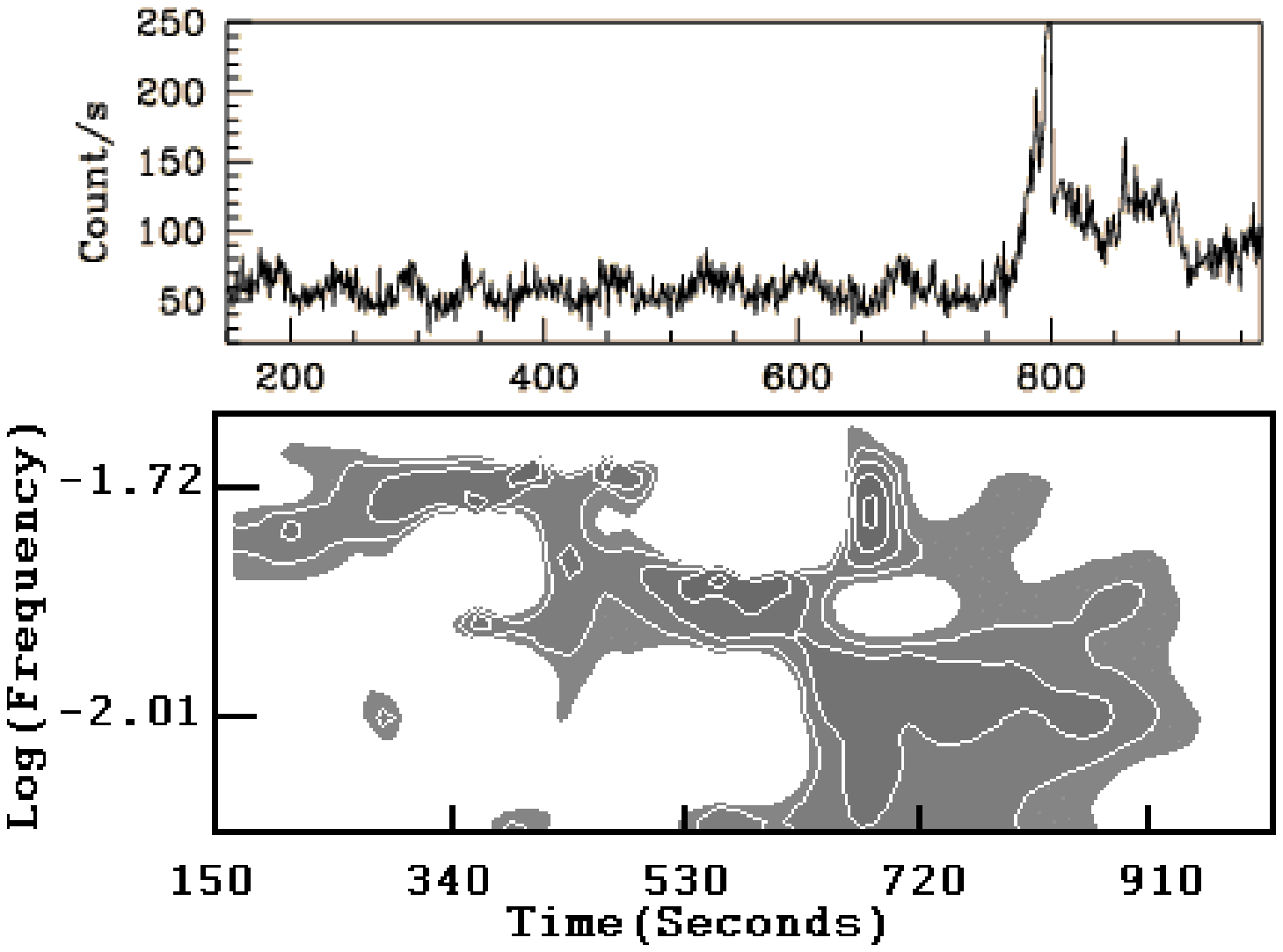} 
\end{figure}

\vfil\eject
\begin{figure}
\vskip 0.0cm
\plotone{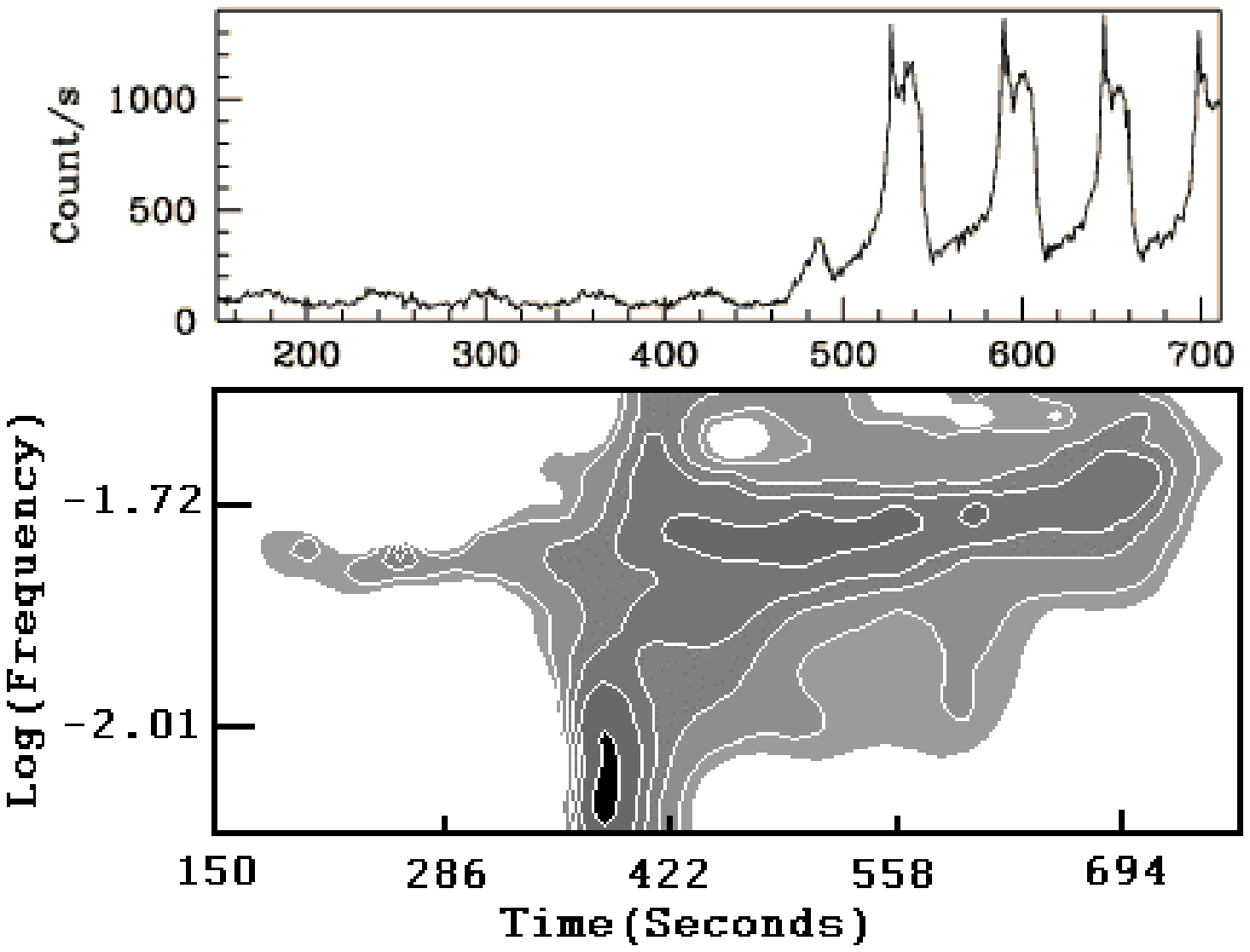}
\end{figure}

\vfil\eject
\begin{figure}
\vskip 0.0cm
\plotone{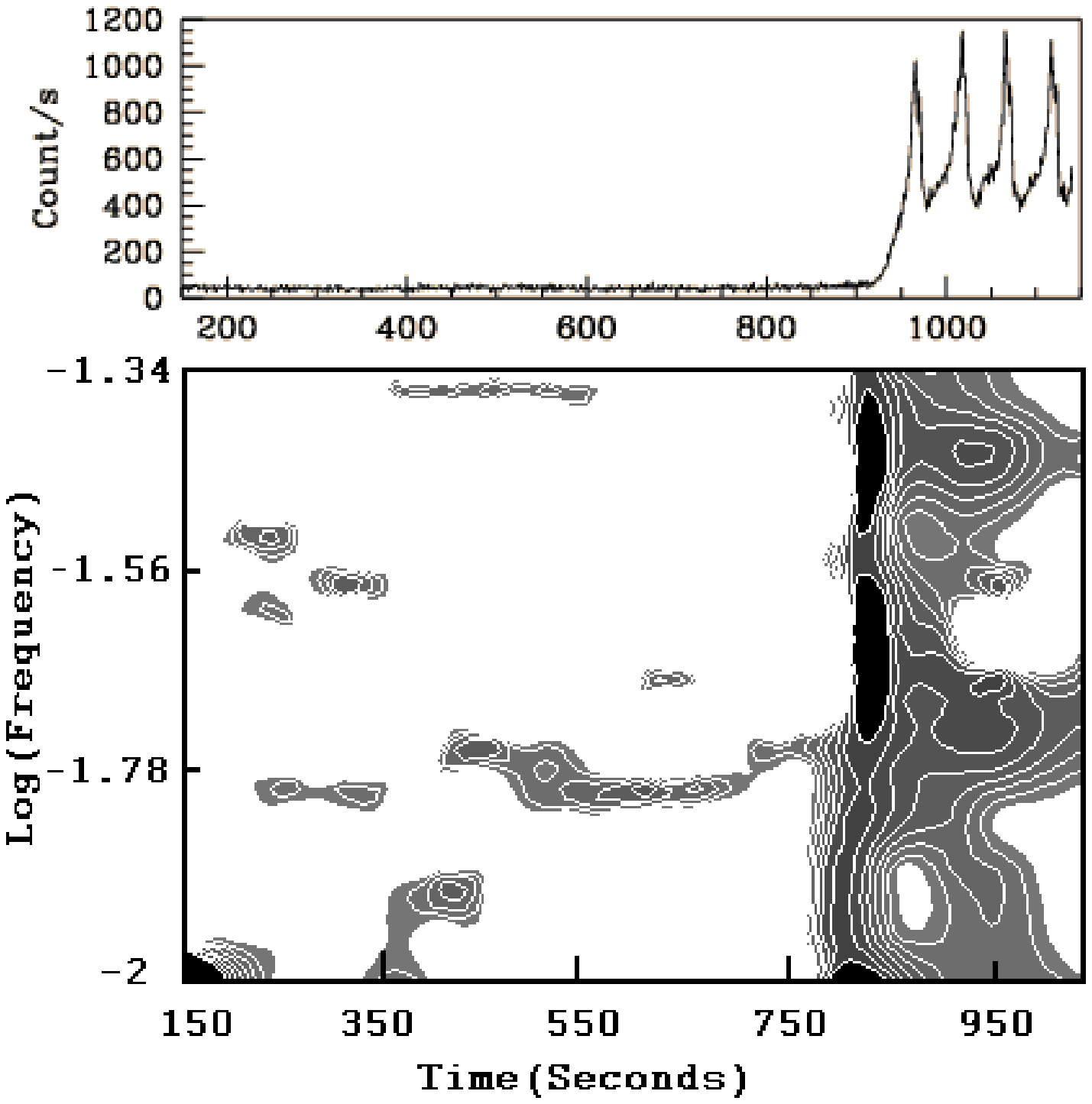}
\end{figure}

\vfil\eject
\vskip 0.0cm
\begin{figure}
\plotone{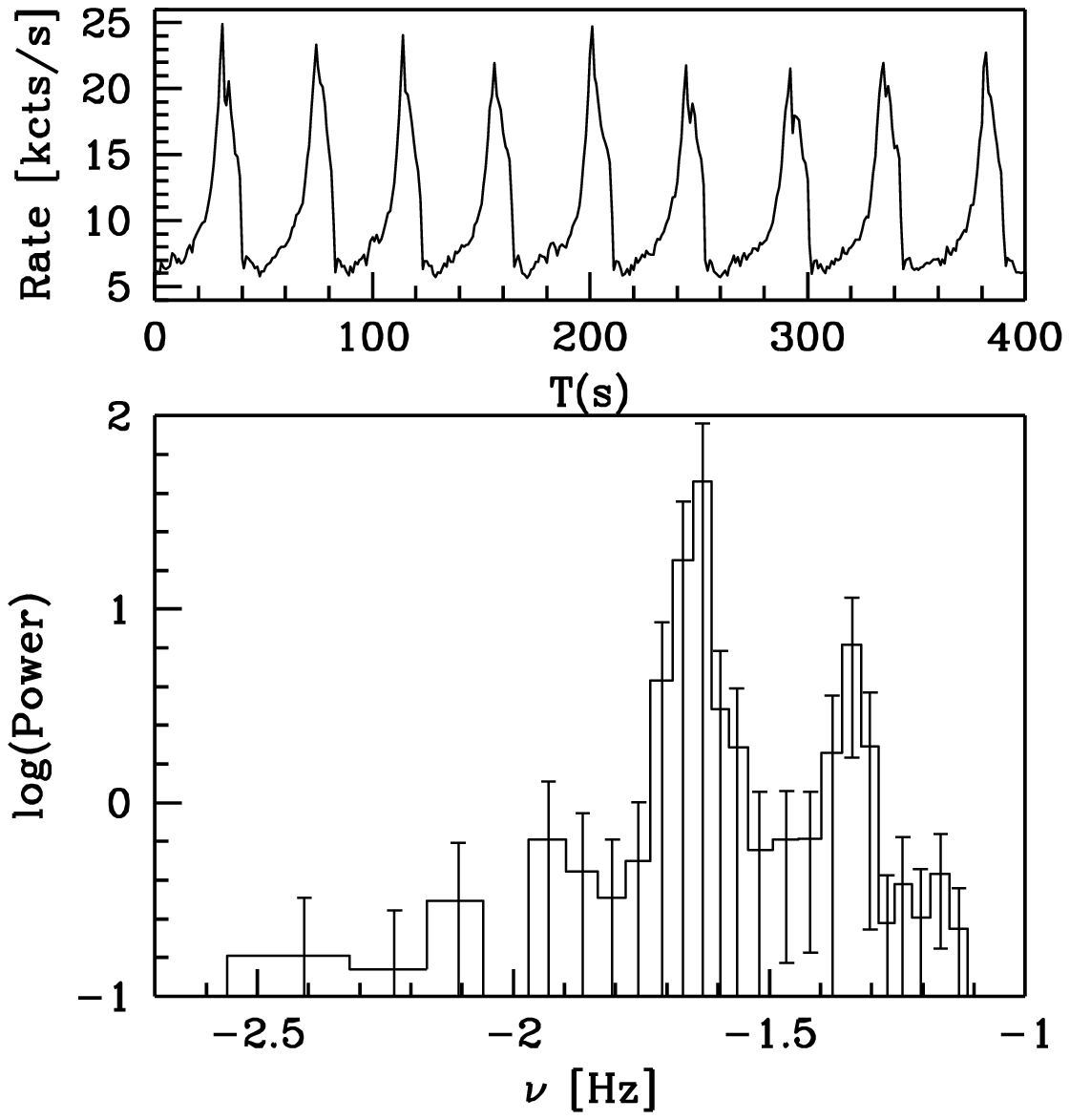}
\end{figure}

\newpage 

\noindent Fig. 1(a-e): Light curves (upper panel) and time-frequency plots (lower panel) of GRS 1915+105
as observed by IXAE in successive five orbits on the 22nd of June, 1997. In (a) the light curve is of
$\kappa$ class while in (b), (c) and part of (d) the light curve is of unknown class. In (d),
a failed attempt for transition was made where a `spike' is seen (as is evidenced from the disparate power-density 
spectrum in the pre- and post-transition period). In (e), a real transition to $\rho$ class is seen.
This is also reflected in the time-frequency plot. 

\noindent Fig. 2: Light curves (upper panel) and time-frequency plots (lower panel) of GRS 1915+105
as observed by IXAE on the 25nd of June, 1997. Upper panel shows the transition from $\chi$ to
$\rho$ class. The lower panel shows the characteristic noise in the power density spectrum at the transition. 

\noindent Fig. 3: RXTE observation of GRS 1915+105 after half an hour of the IXAE observation as
presented in Fig. 2. Upper panel shows that the light curve is still in $\rho$ class. The power 
density spectrum in lower panel shows the low frequency QPO as well close to where IXAE found it.

\end{document}